\documentclass[11pt]{article}

\usepackage{euscript}
\usepackage{amssymb}
\usepackage{amsfonts}
\usepackage{amsbsy}
\usepackage{amsmath}
\usepackage{epsfig}
\usepackage{amsthm}
\usepackage{amscd}
\usepackage{amstext}
\usepackage{verbatim}

\usepackage[pdftex]{hyperref}

\def\ben{\begin{equation}}
\def\een{\end{equation}}
\def\half{{\textstyle{1\over2}}}

 \let\b=\beta \let\g=\gamma \let\d=\delta 
   \let\k=\kappa

\let\w=\omega

\let\pa=\partial
\def\be{\begin{equation}}
\def\ee{\end{equation}}
\def\beq{\begin{equation}}
\def\eeq{\end{equation}}
\def\ba{\begin{array}}
\def\ea{\end{array}}

\def\dalemb#1#2{{\vbox{\hrule height .#2pt
       \hbox{\vrule width.#2pt height#1pt \kern#1pt
               \vrule width.#2pt}
       \hrule height.#2pt}}}

\newcommand{\bea}{\begin{eqnarray}}
\newcommand{\eea}{\end{eqnarray}}

\def\R{{{\Bbb R}}}

\def\Lag{{\mathcal{L}}}

\def\ocal{{\mathcal{O}}}

\begin{document}

\begin{center}

{ \LARGE {\bf Fractionalization of holographic Fermi surfaces}}

\vspace{1cm}

{\large Sean A. Hartnoll$^\flat$ and Liza Huijse$^\sharp$

\vspace{0.7cm}

{\it $^\flat$ Department of Physics, Stanford University, \\
Stanford, CA 94305-4060, USA \\}

\vspace{0.3cm}

{\it $^\sharp$ Department of Physics, Harvard University,\\
Cambridge, MA 02138, USA \\} }

\vspace{1.6cm}

\end{center}

\begin{abstract}

Zero temperature states of matter are holographically described by a spacetime with an asymptotic electric flux.
This flux can be sourced either by explicit charged matter fields in the bulk, by an extremal black hole horizon, or by a combination of the two.
We refer to these as mesonic, fully fractionalized and partially fractionalized phases of matter, respectively.
By coupling a charged fluid of fermions to an asymptotically $AdS_4$ Einstein-Maxwell-dilaton theory, we exhibit quantum
phase transitions between all three of these phases. The onset of fractionalization can be either a first order or continuous
phase transition. In the latter case, at the quantum critical point the theory displays an emergent Lifshitz scaling symmetry in the IR.

\end{abstract}

\pagebreak
\setcounter{page}{1}

\section{Context}

A field theory with a global $U(1)$ symmetry may be placed at a finite charge density $\langle J^t \rangle$.
It is of interest to characterize the resulting possible low energy, zero temperature phases of matter. In this paper
we will consider phases of matter where no symmetries (i.e. neither the global $U(1)$ nor the spacetime
symmetries) are spontaneously broken. Lorentz invariance will of course be broken by the charge density itself.

The holographic correspondence \cite{Maldacena:1997re} gives a set of theories for which this question may be posed
nonperturbatively. The charge density is dually described by an asymptotic electric flux $\int_{\R^2} \star F$ in the holographically dual
spacetime. This electric flux must be sourced in the interior of the spacetime. In the classical gravity limit there are two
qualitatively distinct sources for this flux: charged matter in the bulk or a charged event horizon (see e.g. \cite{Hartnoll:2011fn}).
The sharpness of this distinction is made possible by the fact that the large $N$ limit of theories with gravity duals
necessarily introduces a parametric hierarchy between a few `light' bulk fields and a large number of `black hole microstates', see e.g.
\cite{ElShowk:2011ag}.

Charged matter fields in the bulk correspond to (composite) gauge invariant charged operators in the dual field theory.
At a classical level in the bulk, such fields must be fermionic if they are to contribute to the charge density without condensing or breaking
a spacetime symmetry. It has recently been understood that the charge density carried by such `mesinos' (i.e. fermionic meson-like operators)
retains key features of conventional Fermi liquids. In particular, the sum over the volumes of all Fermi surfaces in the mesino correlators
equals the total charge density carried by the mesinos, in accordance with the Luttinger theorem \cite{Hartnoll:2011dm, Iqbal:2011in, Sachdev:2011ze}. The phenomenology of these Fermi surfaces can be unconventional due to dissipation into a locally quantum critical sector \cite{Lee:2008xf, Liu:2009dm, Cubrovic:2009ye, Faulkner:2009wj, Denef:2009yy, Hartnoll:2009kk, Faulkner:2010da} or due to the existence of many very closely spaced Fermi surfaces \cite{Hartnoll:2010xj, Hartnoll:2011dm, Puletti:2011pr}. Nonetheless, all of these features can be understood without holography as conventional Fermi surfaces coupled to a critical sector in a way that allows the fermion self energy diagrams to be reliably resummed at large $N$ \cite{Faulkner:2010tq, Sachdev:2010um}.

Charge emanating from behind an event horizon, in contrast, appears as a mismatch in the total
Luttinger count of gauge invariant Fermi surfaces of `light' operators \cite{Hartnoll:2011fn}. It has been argued that such a mismatch is a defining characteristic of fractionalized phases of matter, in which the remaining charge density is thought of as being carried by the gauge-charged `quarks'  \cite{Huijse:2011hp}. Indeed, by analogy with the holographic association of deconfined phases with horizons \cite{Witten:1998zw} it is natural to associate flux emanating from a horizon with fractionalized charge carriers \cite{Hartnoll:2011dm, Iqbal:2011in, Hartnoll:2011fn}. The association of charged horizons to fractionalization remains to be made as precise as that of horizons in general to deconfinement, as we currently lack an order parameter for charge fractionalization analogous to the Polyakov loop.

The onset of fractionalization may be an important ingredient in understanding continuous quantum phase transitions in which a Fermi surface undergoes a major re-organization or disappears entirely \cite{FFL1,FFL2,FFL3}. Such transitions may be important in turn for elucidating the
puzzling phenomenology of, among other materials, the heavy fermion compounds and cuprates. However, conventional field theoretic treatments of gauged Fermi surfaces face serious technical challenges, especially in 2+1 dimensions \cite{ssl}.

In this paper we present and explore a zero temperature holographic framework in which a fraction of the bulk electric flux emanates from behind a horizon, while a fraction is sourced by mesinos. The extent of the fractionalization -- i.e. flux from the horizon -- can be dialed using a relevant operator in the dual UV field theory. For large positive couplings, all the charge is fractionalized. At large negative couplings, all the charge is accounted for by mesino Fermi surfaces. For a finite range of intermediate couplings,
\begin{figure}[h]
\begin{center}
\includegraphics[height=220pt]{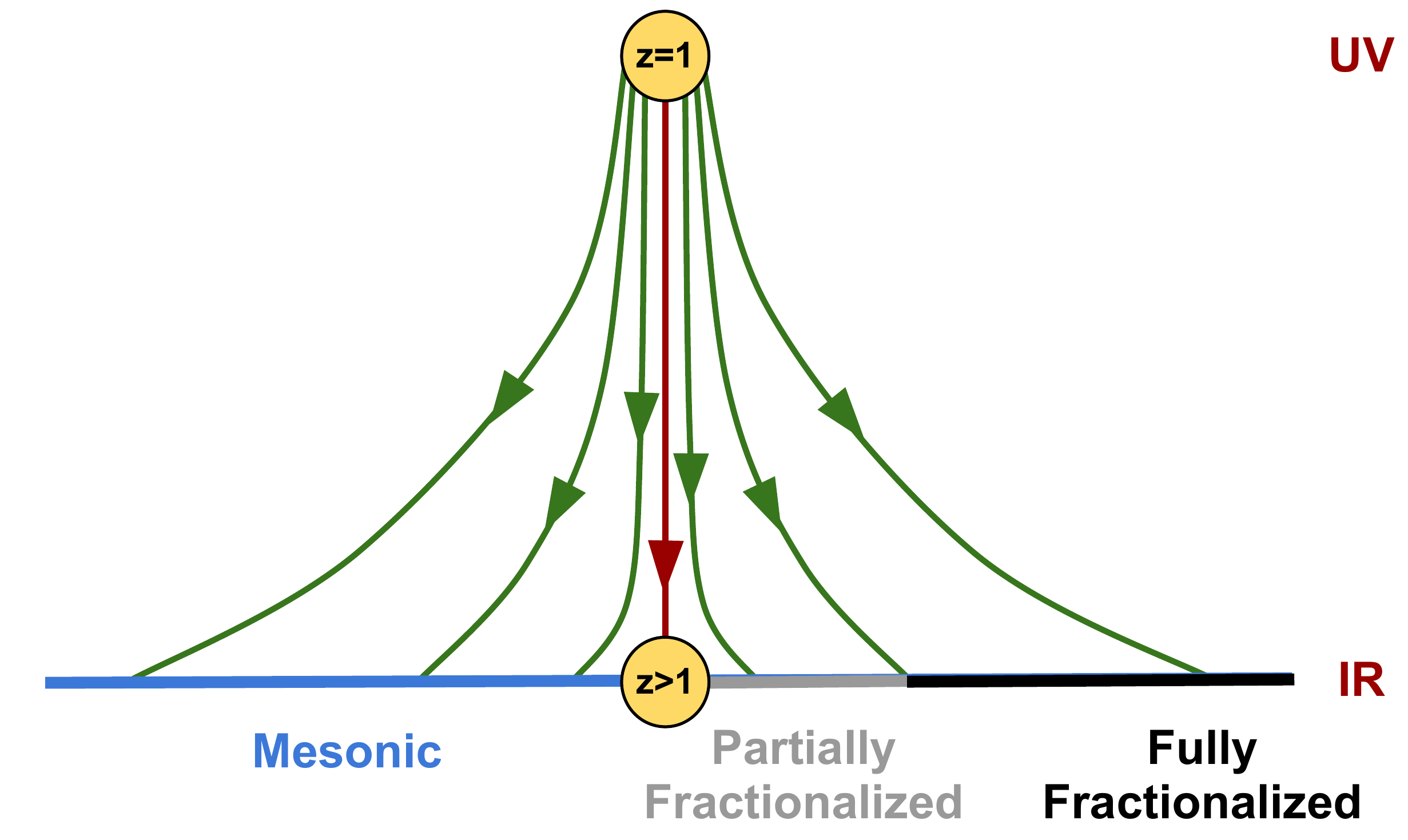}
\caption{Phase diagram. By tuning a relevant coupling in the relativistic UV theory at fixed nonzero chemical potential, the charge density ranges from fully fractionalized to fully mesonic. The fractionalization transition is controlled by an IR fixed point with dynamical critical exponent $z>1$. The IR critical point is sometimes preempted by a first order phase transition. \label{fig:flows}}
\end{center}
\end{figure}
both a charged horizon and mesinos are present. This range of coupling is bounded on the positive end by a continuous phase
transition and on the negative end by a continuous or first order transition, depending on the parameters of the theory.
The continuous transition is controlled by a relevant operator in an emergent IR Lifshitz fixed point. This phase diagram, illustrated in figure \ref{fig:flows} above, is the main result of this paper. The three phases that appear have important similarities with the FL, FL$^*$ and NFL phases discussed in \cite{Huijse:2011hp}.

\section{Bulk theory and equations of motion}

To discuss fractionalization holographically, the minimal bulk ingredients we need are the metric, a Maxwell field and some charged fermionic matter, dual to the mesino operators. The backreaction of the bulk density of fermions on the metric and electromagnetic fields is important \cite{Hartnoll:2009ns}. For this purpose, it is technically convenient, and perhaps natural, to consider the fermions in a limit in which they admit a coarse-grained description as an ideal fluid. In this WKB limit, the fermion wavefunctions become highly localised -- more detailed discussions of the regime of validity of the fluid description can be found in \cite{deBoer:2009wk, Hartnoll:2010gu, Arsiwalla:2010bt, Hartnoll:2011dm}. It is possible to go beyond the fluid limit using the techniques recently considered in \cite{Sachdev:2011ze}, but we shall not do so here.

Let us briefly review the minimal Einstein-Maxwell-charged fluid framework. There are two types of finite density, zero temperature solutions. If the fermion mass is sufficiently large relative to its charge, then the chemical potential in the bulk is insufficient to populate a Fermi sea and the bulk solution is an extremal Reissner-Nordstr\"om black hole. Following the comments in the introduction, we think of this as a fully fractionalized phase. These extremal black holes are stable against pair production of fermions outside the horizon -- because the fermions have more mass than charge (in an appropriate dimensionless sense), pair produced fermions fall through the event horizon rather than being pushed in opposite directions and discharging the black hole. If the fermion is sufficiently light, however, then a Fermi sea is populated in the bulk and, solving the equations of motion, it is found that all the charge resides in the fermion fluid, with no charge emanating from a horizon \cite{Hartnoll:2010gu}. In this case we have a fully mesonic phase that has been dubbed an `electron star'.

Upon heating up the electron star, charge is gradually transferred from the fermion fluid to the horizon. At a critical temperature there is a third order phase transition, above which all the electric flux is sourced by the horizon \cite{Hartnoll:2010ik, Puletti:2010de}. This can be thought of as a very mild cousin of the Oppenheimer-Volkov bound on the mass of neutron stars, which more typically give first order phase transitions \cite{deBoer:2009wk, Arsiwalla:2010bt}. This is the kind of fractionalization transition we will be interested in. However, we would like the transition to occur at zero temperature, where Fermi surfaces are more sharply defined, as a function of some coupling in the field theory. For this we need more ingredients.

A commonly considered generalization of Einstein-Maxwell theory is to add a dilaton field. We will consider the following
Einstein-Maxwell-dilaton-charged fluid action
\be\label{eq:lag}
\Lag = \frac{1}{2 \k^2} \left(R - 2 \left(\nabla \Phi \right)^2 - \frac{V(\Phi)}{L^2} \right)- \frac{Z(\Phi)}{4 e^2} F_{ab} F^{ab} + p(\mu_\text{loc.}) \,.
\ee
Here $p$ is the pressure of the fluid. For our irrotational, zero temperature fluids we can take the local chemical potential in this action to be
\be
\mu_\text{loc.}^2 = g^{ab} A_a A_b \,,
\ee
see for instance \cite{Hartnoll:2010gu}, following \cite{Schutz:1970my}.

Without the charged fluid, this type of action has been considered recently as a model for QCD at finite baryon density \cite{DeWolfe:2010he}, as well as to produce a wide range of extremal near horizon geometries with a view to phenomenological modeling of frequency and temperature scaling laws in condensed matter systems \cite{Taylor:2008tg, Goldstein:2009cv, Charmousis:2010zz}. Various interesting observations about fermionic probes of Einstein-Maxwell-dilaton backgrounds were made in \cite{Gubser:2009qt, Iizuka:2011hg, Rangamani:2011ae}, but will not be directly related to our discussion.
Within the context of QCD, the dilaton is a simple way to incorporate a relevant operator that can mimic the effects of the running QCD coupling and the ultimate onset of confinement, see e.g. \cite{Gubser:2008ny, Gursoy:2010fj}. We shall use the dilaton in a similar way, choosing its mass so that it is dual to a relevant operator.
 
We will study solutions to this theory with metric and Maxwell potential
\bea
ds^2 & = & L^2 \left(- f(r) dt^2 + g(r) dr^2 + \frac{dx^2 + dy^2}{r^2} \right) \,, \\
A & = & \frac{e L}{\k} \, h(r) dt \,.
\eea
Thus in particular the local chemical potential
\be
\mu_\text{loc.} = \frac{e}{\k} \frac{h}{\sqrt{f}} \,.
\ee
The UV boundary will be at $r = 0$, while the IR interior will be at $r \to \infty$. The energy density, charge density and pressure of the fluid are not independent variables, but are rather determined at each point by this local chemical potential. If we introduce the rescaled dimensionless variables
\be
p = \frac{1}{L^2 \k^2} \hat p \,, \qquad \rho = \frac{1}{L^2 \k^2} \hat \rho \,, \qquad \sigma = \frac{1}{e L^2 \k} \hat \sigma \,,
\ee
then the equation of state for the zero temperature Dirac fermion of mass $m$ is
\be\label{eq:eos}
\hat \rho = \hat \b \int_{\hat m}^{\frac{h}{\sqrt{f}}}  \epsilon^2 \sqrt{\epsilon^2 - \hat m^2} d\epsilon \,, \qquad \hat \sigma = \hat \b \int_{\hat m}^{\frac{h}{\sqrt{f}}}  \epsilon \sqrt{\epsilon^2 - \hat m^2} d\epsilon \,, \qquad - \hat p = \hat \rho - \frac{h}{\sqrt{f}} \hat \sigma \,, 
\ee
where
\be
\hat \beta = \frac{e^4 L^2}{\k^2} \frac{1}{\pi^2} \,, \qquad \hat m^2 = \frac{\k^2}{e^2} m^2 \,.
\ee
Note that these local thermodynamic quantities obey (as they must)
\be
\hat p\,{}' =  \left(\frac{h}{\sqrt{f}} \right)' \hat \sigma \,.
\ee
A more leisurely exposition of the above may be found in \cite{Hartnoll:2010gu}.

On the ansatz just described, the equations of motion following from our action (\ref{eq:lag}) are found to be
\bea
\frac{1}{r} \left(\frac{f'}{f} + \frac{g'}{g} + \frac{4}{r} \right) + \frac{g h}{\sqrt{f}} \, \hat \sigma + 2 \Phi'^2 & = & 0 \,, \label{eq:ff} \\
\frac{1}{r} \left(\frac{f'}{f} - \frac{1}{r} \right) + g \left( \hat p - \half V(\Phi) \right) - \frac{Z(\Phi) h'^2}{2 f} + \Phi'^2 & = & 0 \,, \\
- \Phi'' + \frac{1}{2} \left(- \frac{f'}{f} + \frac{g'}{g} + \frac{4}{r} \right) \Phi' + \frac{g V'(\Phi)}{4} - \frac{Z'(\Phi) h'^2}{4 f} & = & 0 \,,  \label{eq:dil} \\
\frac{d}{dr} \left(\frac{Z(\Phi) h'}{r^2 \sqrt{f g}}\right) - \frac{\sqrt{g}}{r^2} \hat \sigma & = & 0 \,. \label{eq:flux}
\eea
These equations are seen to imply the following additional conservation equation, cf. \cite{DeWolfe:2010he,Hartnoll:2010gu},
\be\label{eq:magic}
\frac{d}{dr} \left( \frac{2 r^2 Z(\Phi) h h'- (r^2 f)'}{r^4 \sqrt{fg}} \right) = 0 \,.
\ee
Such additional conservation equations are important for the first law of thermodynamics to be satisfied, as they connect data at the horizon and at the asymptotic boundary. For the zero temperature spacetimes to be considered in this paper, this charge will in fact vanish, so that
\be
2 r^2 Z(\Phi) h h'- (r^2 f)' = 0 \,.
\ee
This can be checked explicitly on the all the near horizon geometries we present below.

\section{Critical points and near horizon solutions}

\subsection{Asymptotic $AdS_4$ solution and thermodynamics}
\label{sec:ads}

We will choose the dilaton potential such that there is an $AdS_4$ solution to our equations of motion, with no fermion fluid. We will use this solution to set the UV boundary conditions on our theory. The general near boundary, $r \to 0$, expansion will be characterized by several constants. These are the boundary speed of light $c$, the energy density $\hat E$, the source and expectation values $\phi_0$ and $\langle \ocal \rangle$ of the operator $\ocal$ dual to the bulk dilaton $\Phi$, and finally the charge density $\hat Q$ and boundary chemical potential $\hat \mu$. The various hats indicate that we are dealing with quantities that have been rescaled by factors of $e ,L, \k$. At this point we will assume, for ease of computation, that the potential has the form near $\Phi = 0$ of
\be\label{eq:Vexpand}
V(\Phi) = - 6 - 4 \Phi^2 + \ocal(\Phi^4) \,,
\ee
so that the dilaton has mass squared of $-2/L^2$. This is above the $AdS_4$ Breitenlohner-Freedman bound; hence the dual operator $\ocal$ has scaling dimension $\Delta = 2$ in the UV theory. We will also set $Z(0)=1$ without loss of generality. The near boundary expansion then takes the form
\bea
f & = & \frac{c^2}{r^2} \left(1 - \left( \hat E + {\textstyle \frac{1}{3}} \phi_0 \langle \ocal \rangle\right) r^3 + \cdots \right) \,, \\
g & = & \frac{1}{r^2} \left(1 - \phi_0^2 r^2 + \left(\hat E - \phi_0 \langle \ocal \rangle \right) r^3 + \cdots  \right) \,, \\
h & = & c \left( \hat \mu - \hat Q r +  \cdots \right) \,, \\
\Phi & = & \phi_0 r + \frac{\langle \ocal \rangle}{2} r^2 + \cdots \,.
\eea
Thus the theory is asymptotically characterized by three sources $\{c,\phi_0,\hat \mu\}$ and three expectation values $\{\hat E, \langle \ocal \rangle,\hat Q\}$.

Using the equations of motion, the bulk Lagrangian (\ref{eq:lag}) is found to become a total derivative on shell
\be
\left. \sqrt{-g} \Lag \right|_\text{on-shell} = \frac{L^2}{\k^2} \frac{d}{dr} \left( \frac{2 Z(\Phi) h h'  - f'}{2 r^2 \sqrt{f g}} \right) = \frac{L^2}{\k^2} \frac{d}{dr} \frac{f}{r^3 \sqrt{fg}}\,.
\ee
In the second equality we used the conservation law (\ref{eq:magic}). To obtain the free energy from the on shell action, we need to add the usual boundary counterterms, see e.g. \cite{Hartnoll:2009sz} and references therein,
\be
\Lag_\text{c.t.} = \frac{1}{\k^2} \left(K - \frac{2}{L} \right) - \frac{1}{\k^2 L} \Phi^2 \,.
\ee
Here $K = \gamma^{\mu \nu}\nabla_\mu n_\nu$ with $\g$ the induced metric on the asymptotic boundary and $n$ an outward pointing unit normal.
The free energy density is now found to be
\bea
\frac{\hat \Omega}{V} & = & - \frac{1}{c} \lim_{\epsilon \to 0} \left(\int_\epsilon  \sqrt{-g} \Lag dr + \left. \sqrt{-\g}\Lag_\text{c.t.} \right|_{r=\epsilon}  \right) \\
& = & \hat E - \hat \mu \hat Q \,.\label{eq:first}
\eea
Here we have used the above expansions and imposed that we are at zero temperature and hence there is no contribution from the endpoint of the radial integral in the interior of the spacetime. Evaluating the conservation law (\ref{eq:magic}) in the asymptotic region furthermore implies the relationship
\be\label{eq:extra}
\frac{3}{2} \hat E - \hat \mu \hat Q + \frac{1}{2} \phi_0 \langle \ocal \rangle  = 0 \,.
\ee
This expression follows from tracelessness of the energy momentum tensor for a conformally invariant theory, $\hat E = 2 \hat P$, together with the effect of the relevant deformation $\phi_0$. Consider changing the volume at fixed chemical potential and zero temperature.
Differentiate the free energy (\ref{eq:first}) and simplify using the first law of thermodynamics, $d(\hat E V) = - \hat P dV - \langle \ocal \rangle d(\phi_0 V) + \hat \mu \, d (\hat Q V)$. Extensivity of the free energy allows us to integrate back again to give $\hat \Omega/V = - \hat P - \langle \ocal \rangle d(\phi_0 V)/dV = - \hat P - \frac{1}{2} \phi_0 \langle \ocal \rangle$. In the second step we used the fact that $\phi_0$ has scaling dimension one. Using the tracefree condition, we obtain (\ref{eq:extra}).

\subsection{Lifshitz fixed point}
\label{sec:lif}

In the presence of the charged fluid, our system admits a second scale invariant solution. This will be a Lifshitz spacetime \cite{Kachru:2008yh} characterized by a dynamical critical exponent $z$. The dilaton will be constant, and therefore the physics governing these solutions is essentially the same as that of the IR geometry of electron stars without the additional dilaton field \cite{Hartnoll:2009ns, Hartnoll:2010gu}. The difference will be the presence of a relevant perturbation of the Lifshitz fixed point. The system must therefore be tuned in the UV to hit this IR Lifshitz solution. We will discuss the renormalization group trajectories below, as well as the nature of the continuous phase transition that occurs when the system is driven through the Lifshitz fixed point.

The Lifshitz solution takes the form
\be
ds^2 = L^2 \left(- \frac{dt^2}{r^{2z}} + g_L \frac{dr^2}{r^2} + \frac{dx^2 + dy^2}{r^2} \right) \,, \qquad A = h_L \frac{eL}{\k} \frac{dt}{r^z} \,, \qquad \Phi = \phi_L \,.
\ee
So there are four constants to solve for: $\{z, g_L, h_L, \phi_L\}$. The dilaton equation of motion (\ref{eq:dil}) implies that the field must be at the minimum of an effective potential receiving contributions from the potential and also from the electric flux
\be
V_\text{eff.}'(\phi_L) = g_L V'(\phi_L) - z^2 h_L^2 Z'(\phi_L) = 0 \,.
\ee
In general this equation may have multiple solutions.
Meanwhile, the Gauss law (\ref{eq:flux}) selfconsistently relates the potential $h_L$ to the constant charge density $\hat \sigma(h_L)$
\be
2 z h_L Z(\phi_L) = g_L \hat \sigma(h_L) \,.
\ee
The conservation law (\ref{eq:magic}) gives a simple expression for the dynamical critical exponent
\be
z = \frac{1}{1 - h_L^2 Z(\phi_L)} \,.
\ee
Thus in particular $z > 1$ if $Z(\Phi)$ is positive. The remaining equation of motion is then seen to give, after simplification using the above expressions,
\be
(1+z)(2+z) = 2 g_L \hat p(h_L) - g_L V(\phi_L) \,.
\ee
For a given theory with $V(\Phi)$ and $Z(\Phi)$, we can use these equations to eliminate $z$ and $g_L$ and obtain coupled equations for $\{h_L,\phi_L\}$ that must be solved numerically.

For our Lifshitz spacetimes, the conserved electric flux $\int_{\R^2} \star \left[ Z(\Phi) F \right] \sim r^{-2} \to 0$ in the interior as $r \to \infty$. Thus there is no flux emanating from the Lifshitz `horizon', and the Lifshitz IR describes a fully mesonic configuration.

At this point it is convenient to fix our model. For the remainder of the paper,
we choose the following simple form for the scalar potential and coupling to the Maxwell term
\be\label{eq:ZV}
Z(\Phi) = e^{2 \, \Phi/\sqrt{3}} \,, \qquad V(\Phi) = - 6 \cosh \left({\textstyle{2 \, \Phi/\sqrt{3}}} \right) \,.
\ee
This form was largely chosen for convenience of numerical implementation. The potential satisfies our assumed expansion (\ref{eq:Vexpand}) at small fields, so that the operator $\ocal$ dual to the dilaton is relevant in the UV with dimension $\Delta = 2$. The fact that $Z$ is not symmetric in $\Phi$ implies that $\Phi = 0$ will not be a solution to the equations of motion once there is electric flux, even when the dual operator $\ocal$ is not sourced explicitly. Many features we are interested in will not depend on the details of $Z$ and $V$, we shall comment on this as appropriate below.\footnote{A positive energy theorem for the Einstein-dilaton sector of our theory is equivalent to the existence of a superpotential $W$ \cite{Townsend:1984iu}, which with our normalization must satisfy $V = 2 (\pa_\Phi W)^2 - 6 W^2$. While we have not found $W$ analytically for our potential (\ref{eq:ZV}), we have checked numerically that it exists and has asymptotic, large $\Phi$, behavior $W \sim \cosh \sqrt{3} \Phi$.}

With the expression (\ref{eq:ZV}) it is straightforward to solve the algebraic equations for $\{z, g_L, h_L, \phi_L\}$ for a given $\{\hat \beta, \hat m\}$, characterizing the fermion fluid. The Lifshitz solution can only exist if the local chemical potential is large enough to populate the Fermi sea, which requires
\be
\hat m < \hat m_\text{max.} = h_L(\hat \beta, \hat m_\text{max.}) \,.
\ee
The above equality may be numerically solved to find $\hat m_\text{max.}$.
The following figure \ref{fig:mmax} shows $\hat m_\text{max.}$ as a function of $\hat \beta$.
\begin{figure}[h]
\begin{center}
\includegraphics[height=180pt]{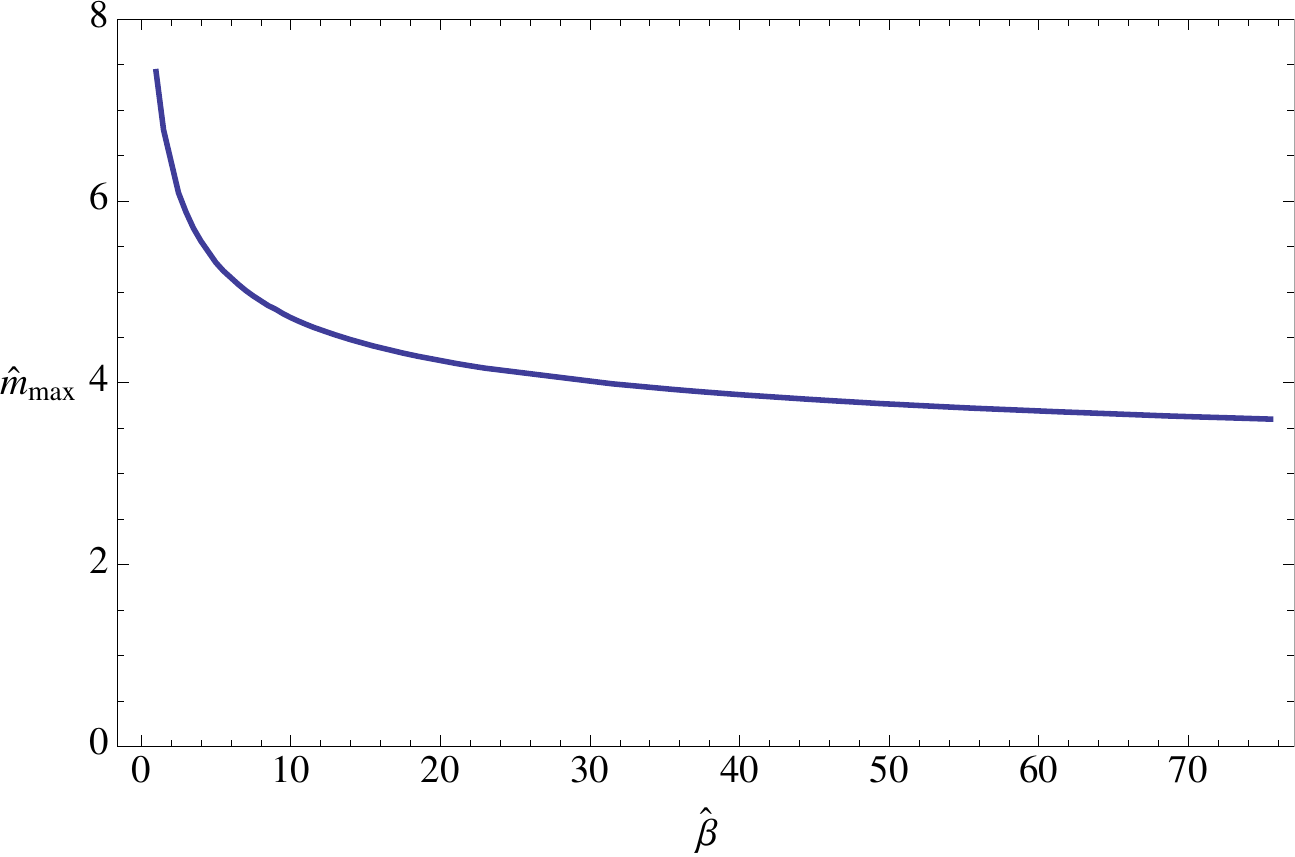}
\caption{Maximal mass $\hat m_\text{max.}$ as a function of $\hat \beta$. The Lifshitz solution requires $\hat m < \hat m_\text{max.}$. \label{fig:mmax}}
\end{center}
\end{figure}

Given the Lifshitz solution, we need to characterize the deformations away from the fixed point. To this end, write
\bea
h = \frac{h_L}{r^z} \left(1 + \d h \, r^M \right) \,, \qquad g = \frac{g_L}{r^2} \left(1 + \d g \, r^M \right) \,, \\
f = \frac{1}{r^{2z}} \left(1 + \d f \, r^M \right) \,, \qquad \Phi = \phi_L \left(1 + \d \phi \, r^M \right) \,.
\eea
Because the total order of our equations of motion (\ref{eq:ff}) -- (\ref{eq:flux}) is six, we can expect to obtain six different possible exponents $M$. These exponents will consist of three pairs of sources with momentum scaling dimensions $[g_\ocal] = M_-$ and expectation values with scaling dimensions $[\langle \ocal \rangle]  = M_+$ satisfying
\be\label{eq:scalingdims}
M_+ + M_- = 2 + z \,.
\ee
This relation comes from the fact that each pair of modes describes the insertion of an operator $\int dt d^2x \, g_\ocal \, \ocal$ at the Lifshitz fixed point.
Here $\ocal$ represents any one of the three operators in the Lifshitz IR theory, not the operator dual to the dilaton.

A key fact we will wish to characterize is whether the operators that can perturb the Lifshitz fixed point are relevant or irrelevant. There is a universal deformation with $[g_\ocal] = 0$ and associated expectation value with dimension $2+z$ that corresponds to placing the theory at a finite temperature (cf. for instance \cite{Gubser:2009cg, Goldstein:2009cv, Hartnoll:2010gu}). We will not want to source this deformation. We then find that of the remaining two operators, one is irrelevant and one is relevant. The irrelevant operator has $[g_\ocal] < 0$, and hence becomes more important towards the UV boundary at $r \to 0$, while the relevant operator has $[g_\ocal] > 0$.

\begin{figure}[h]
\begin{center}
\includegraphics[height=210pt]{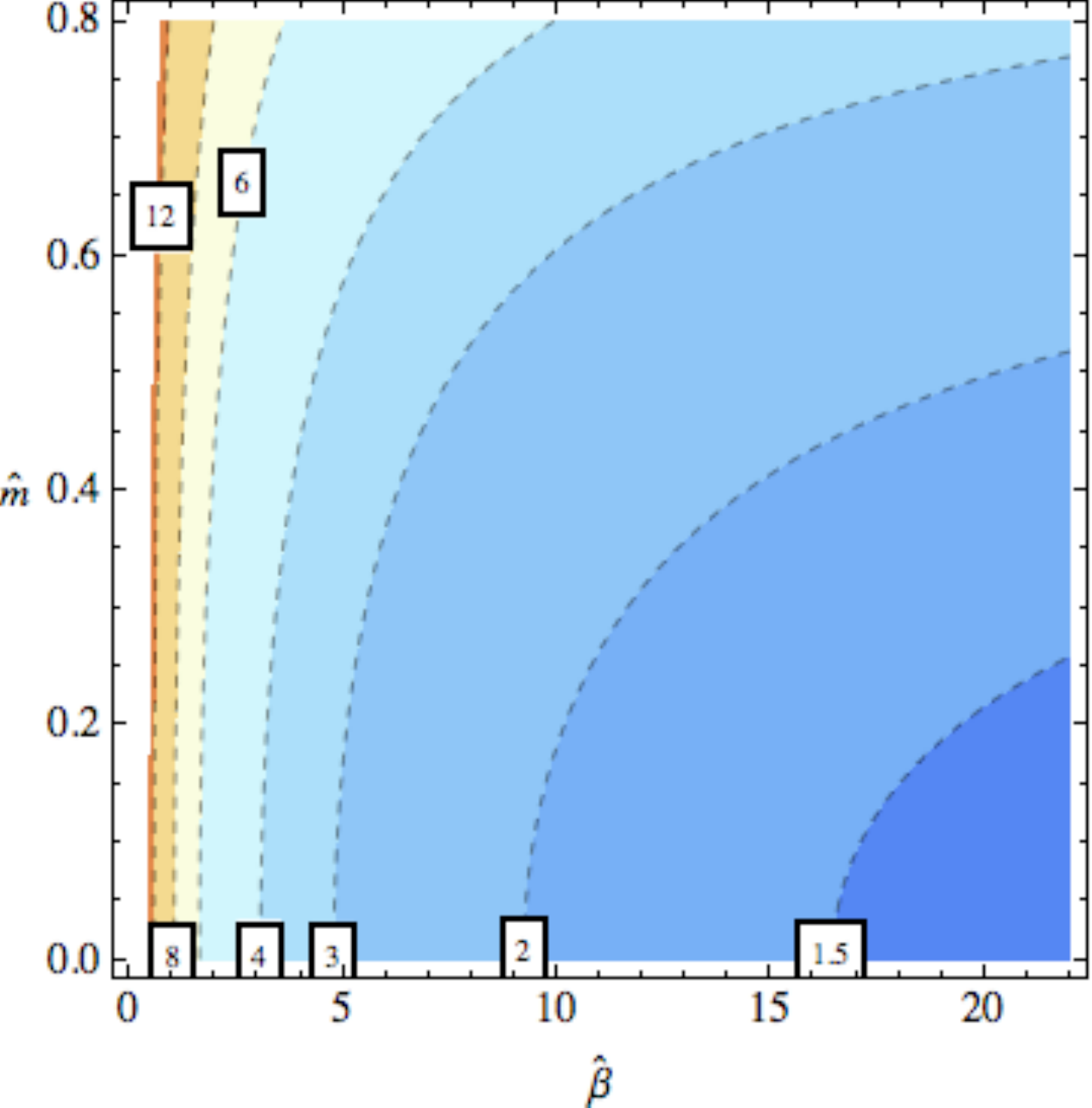}\hspace{0.5cm}\includegraphics[height=210pt]{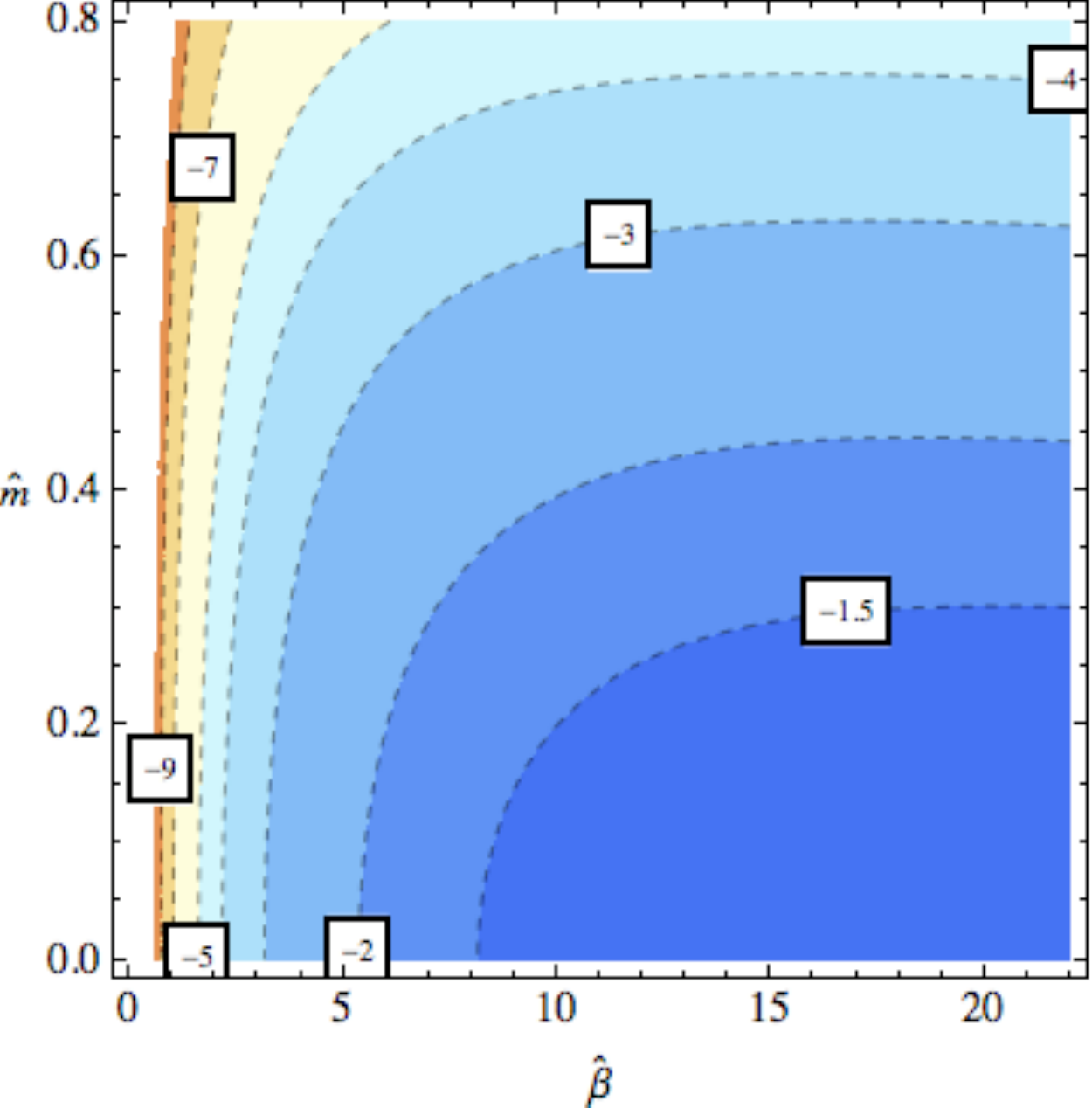}
\caption{Contour plot of the dynamical critical exponent $z$ (left) and dimension $[g_\ocal]$ of the irrelevant operator (right) as a function of $\{\hat \b, \hat m \}$. \label{fig:zcontour}}
\end{center}
\end{figure}

Consider the irrelevant deformation first. We will find below, when we consider full RG flows, that the irrelevant operator allows us to flow up from the Lifshitz fixed point to the relativistic UV fixed point. See figure \ref{fig:flows} above. In figure \ref{fig:zcontour} we show the dimension of the irrelevant operator, and also the value of the dynamical critical exponent, for a range of allowed values of $\{\hat \b, \hat m \}$.
The dynamical critical exponent and dimension of the irrelevant operator will determine the scaling of quantities as the Lifshitz point is approached from the UV. In particular, they will determine the low temperature scalings of observables. It is of interest to heat up the phase diagram in figure \ref{fig:flows}, to describe the thermodynamic crossovers around the critical point, along the lines of the recent interesting work \cite{D'Hoker:2010rz}.

The presence of the relevant deformation means that we will need to tune the UV theory if we want to hit the Lifshitz fixed point in the IR. Indeed, we have a dimensionless ratio of relevant couplings that can be tuned in the UV: $\phi_0/\hat \mu$ from section \ref{sec:ads} above. We can expect that a critical value of this ratio will correspond to the critical red line in figure \ref{fig:flows} that flows between the two fixed points. A little away from the critical ratio, as the flow runs close to the IR Lifshitz fixed point, it will pick up the relevant coupling in the IR and flow away from the Lifshitz fixed point. We can therefore anticipate that the Lifshitz fixed point mediates a transition between two distinct phases of matter. Indeed we will find below that these correspond to mesonic and fractionalized phases. One important subtlety here is that the dimension of the relevant operator turns out to be complex for a range of values of $\{\hat \b, \hat m \}$, as we show in figure \ref{fig:reimcontour} below. This phenomenon has been observed previously about Lifshitz fixed points, e.g. in \cite{Gubser:2009cg}. The two powers $M_\pm$ are complex conjugates of each other. In this case (\ref{eq:scalingdims}) necessarily implies $\text{Re}\, [g_\ocal] = (2+z)/2$.
\begin{figure}[h!]
\begin{center}
\includegraphics[height=210pt]{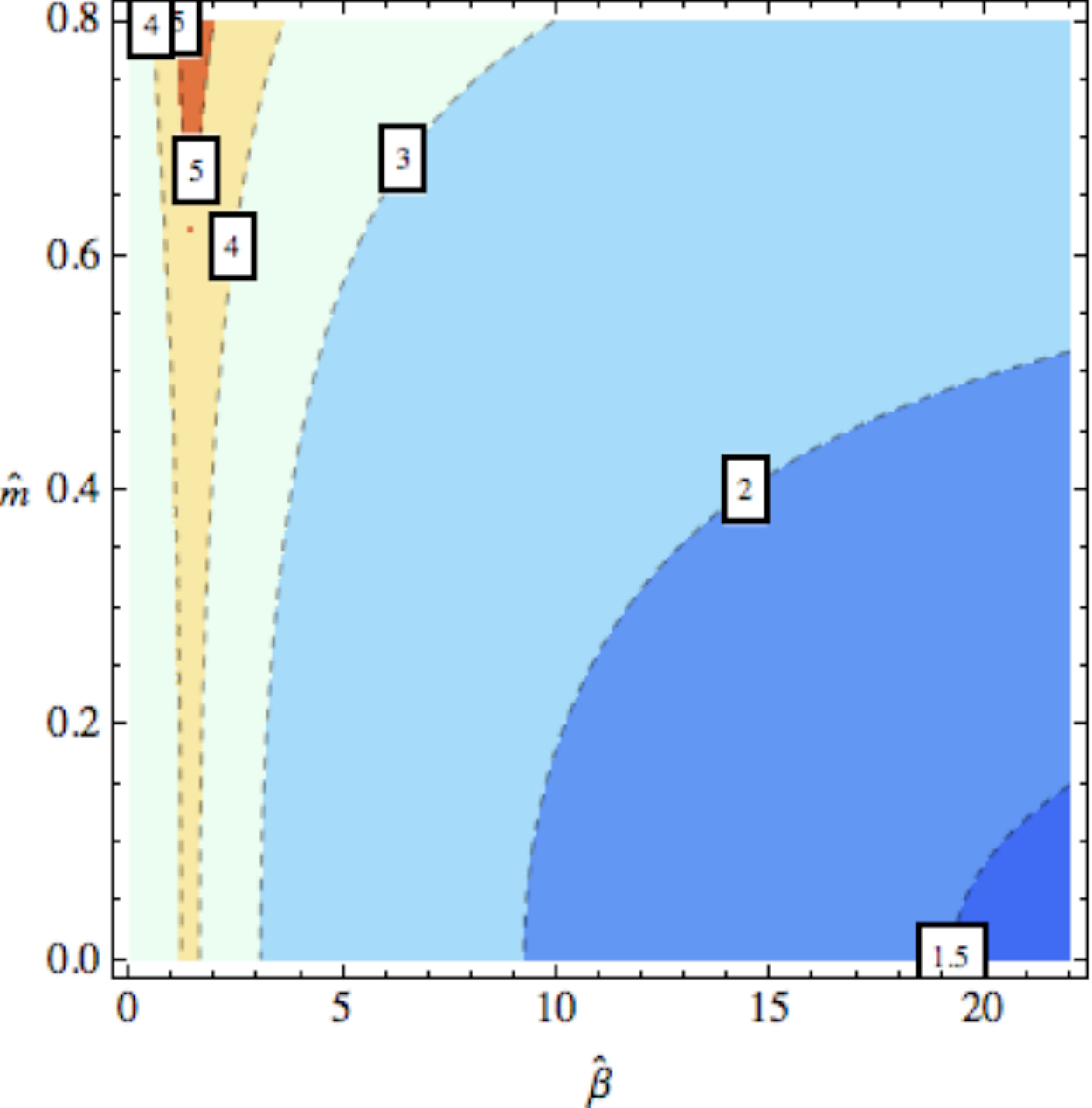}\hspace{0.5cm}\includegraphics[height=210pt]{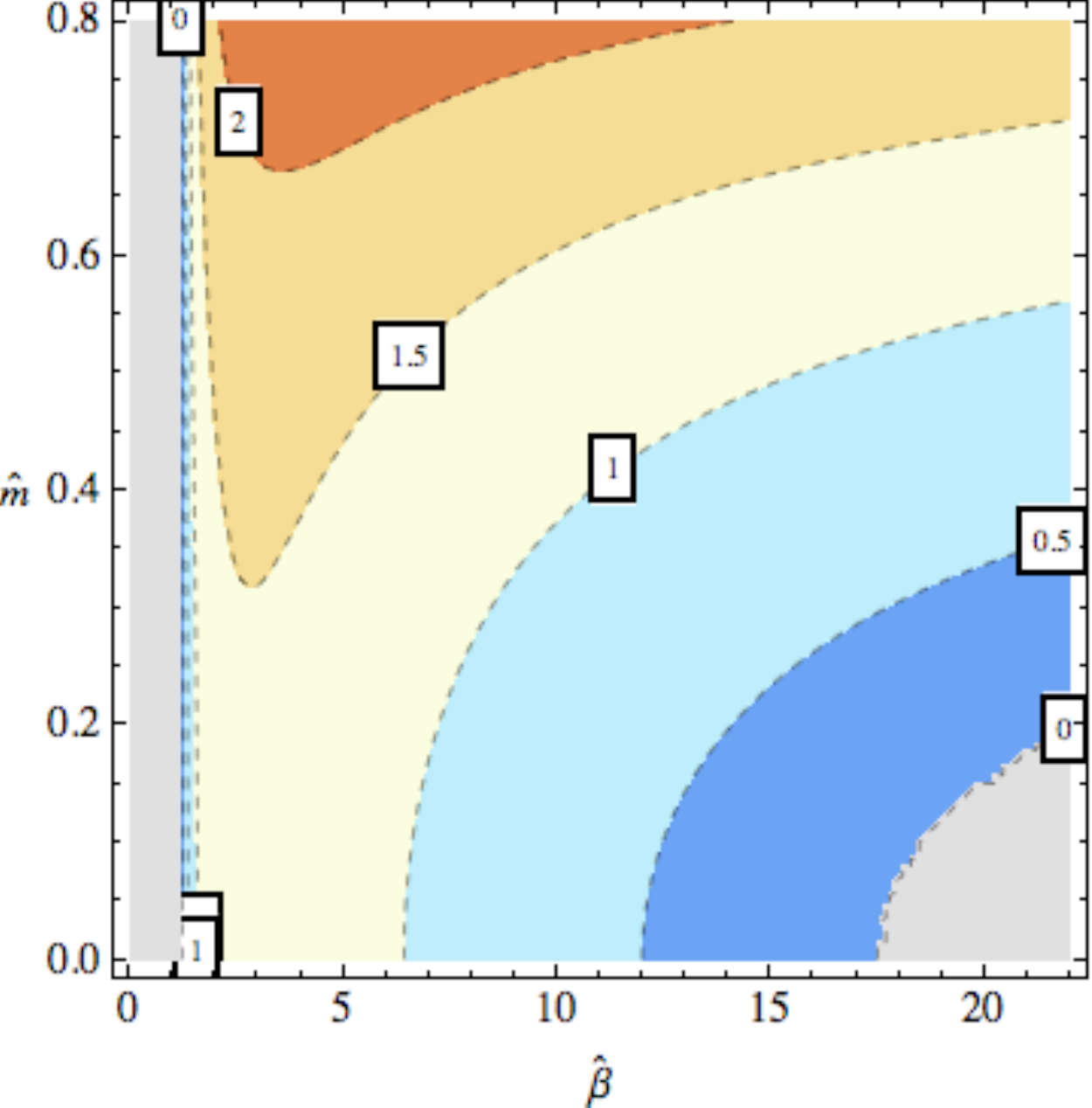}
\caption{Contour plot of the real (left) and imaginary (right) parts of the scaling dimension $[g_\ocal]$ of the relevant operator as a function of $\{\hat \b, \hat m \}$. A nonzero imaginary part indicates an instability of the Lifshitz solution, leading to a first order phase transistion. \label{fig:reimcontour}}
\end{center}
\end{figure}
Complex scaling exponents are known to indicate the presence of an instability, see \cite{Denef:2009tp} and \cite{Edalati:2011yv} for $AdS_2 \times \R^2$ and Lifshitz examples, 
respectively. Indeed, we will find below that when the scaling dimensions are complex, the Lifshitz solution is unstable and the continuous phase transition is preempted by a first order transition between mesonic and partially fractionalized phases.

At the Lifshitz fixed point we have just described, the dilaton was stabilized in the IR at the value $\Phi = \phi_L$. In the following two subsections we will see that the two phases on either side of the Lifshitz fixed point are characterized by runaway IR behaviors in opposite directions: $\Phi \to \pm \infty$. This distinction is possible because we chose $Z$ in (\ref{eq:ZV}) not to be symmetric under $\Phi \leftrightarrow - \Phi$.

\subsection{Near horizon geometries with flux (fractionalized)}
\label{sec:a}

In this subsection we look for IR geometries, i.e. $r \to \infty$, such that the dilaton becomes large and positive. It follows from (\ref{eq:ZV}) that $Z \to \infty$, and hence the effective Maxwell coupling is becoming small. The solution can be determined as a series expansion together with a perturbation that will
allow the flow to be integrated up to the UV:
\bea
f & = &  \frac{1}{r^6}\left(1 + \sum_{n=1}^\infty \frac{f_n}{r^{4n}} + \delta f \, r^{N} \right) \,, \label{eq:ffa} \\
g & = & \frac{16}{3} \frac{1}{r^4} \left(1 + \sum_{n=1}^\infty \frac{g_n}{r^{4n}} + \delta g \, r^{N} \right)  \,, \\
h & = & \frac{1}{\sqrt{2}}\frac{1}{r^4} \left(1 + \sum_{n=1}^\infty \frac{h_n}{r^{4n}} + \delta h \, r^{N}\right)  \,, \\
\Phi & = & \sqrt{3} \log r + \sum_{n=1}^\infty \frac{p_n}{r^{4n}}  + \delta \phi \, r^{N}\,. \label{eq:pp}
\eea
The coefficients $\{f_n, g_n, h_n, p_n\}$ are uniquely determined by the equations of motion, while the perturbation $\{\delta f, \d g, \d h, \d \phi \}$ has
an overall free magnitude. This irrelevant pertubation has
\be
N = 2 - 2\sqrt{19/3} < 0 \,.
\ee
We can expect that different magnitudes for the perturbation will map onto different values of the dimensionless ratio of relevant couplings $\phi_0/\hat \mu$ in the UV theory.

From the IR expansion above we have that at leading order the local chemical potential $\hat \mu_\text{loc.} \sim r^{-1} \to 0$. It follows that for any nonzero fermion mass $\hat m$, there will be no fluid in the far IR of the spacetime. Because $\hat \mu_\text{loc.}$ is increasing as we go towards the UV, it may be the case that there is a fluid of fermions in an intermediate region of the spacetime, similarly to the finite temperature spacetimes considered in \cite{Hartnoll:2010ik, Puletti:2010de}. We will see this explicitly later when we perform a full integration of the equations.

The conserved electric flux, in contrast, tends to a constant as $r\to \infty$, leading to a nonzero flux emanating from the null singularity in the far IR: ${\displaystyle{ \lim_{r \to \infty}}} \int_{\R^2} \star \left[ Z(\Phi) F \right] \sim \text{const.}$. The emergence of flux from behind a `horizon' indicates that these IR geometries describe phases of matter that are at least partially fractionalized.

The IR solution (\ref{eq:ffa}) -- (\ref{eq:pp}) is not scale invariant as $r\to \infty$. This is manifested in the fact that $g \, dr^2$ scales, as well as the logarithmic dependence of the dilaton. The curvature invariants are not constant and indeed become large in the IR. Lack of scale invariance implies that we have a set a dimensionful scale to unity in (\ref{eq:ffa}) -- (\ref{eq:pp}). We shall only compute dimensionless ratios of physical quantities in the UV, and therefore this scale will drop out. While the effective `string coupling' is becoming small in the limit $Z \to \infty$, the divergence of curvature invariants indicates that higher derivative corrections eventually become important at the lowest energy scales. 

The IR geometry we have just presented, without a charged fluid, is a special case of the families of IR solutions discussed in, e.g. \cite{Charmousis:2010zz, Gubser:2009qt, Iizuka:2011hg}. Those papers also considered the effects of a small temperature on the geometry, by finding near-extremal black hole solutions, and computed the temperature dependence of the specific heat. Specialized to our case, their formulae show that $c_s \sim T$ at low temperatures. A power law dependence of the specific heat indicates that this phase of the system contains gapless excitations.

\subsection{Near horizon geometries without flux (mesonic)}
\label{sec:b}

In this subsection we look for IR geometries such that the dilaton becomes large and negative. It now follows from (\ref{eq:ZV}) that $Z \to 0$, and hence the effective Maxwell coupling is becoming large. We found the following IR series expansion and irrelevant perturbation in this case:
\bea
f & = &  \frac{1}{r^2}\left(1 + \sum_{n=1}^\infty \frac{f_n}{r^{2n/3}} + \delta f \, r^{P} \right) \,, \label{eq:f1} \\
g & = & \frac{16}{9} \frac{1}{r^{8/3}} \left(1 + \sum_{n=1}^\infty \frac{g_n}{r^{2n/3}} + \delta g \, r^{P} \right)  \,, \\
h & = & \frac{h_0}{r} \left(1 + \sum_{n=1}^\infty \frac{h_n}{r^{2n/3}} + \delta h \, r^{P-2/3}\right)  \,, \\
\Phi & = &  - \frac{\log r}{\sqrt{3}} + \sum_{n=1}^\infty \frac{p_n}{r^{2n/3}}  + \delta \phi \, r^{P}\,. \label{eq:p1}
\eea
Here $\frac{21}{16} h_0 = \hat \sigma(h_0)$, there is a nonvanishing charge density of fluid in this case, and the $\{f_n, g_n, h_n, p_n\}$ are similarly uniquely determined by the equations of motion. As previously, the perturbation $\{\delta f, \d g, \d h, \d \phi \}$ has a free overall magnitude, and generates an RG flow to the relativistic UV fixed point. The exponent
\be
P = 1 - \frac{2}{3} \sqrt{1 + \frac{63 \, h_0^2}{4 (h_0^2 - \hat m^2)}} < 0 \,.
\ee

In contrast to the previous subsection, here we see that $\hat \mu_\text{loc.} \to h_0 > \hat m$, so that a fermion fluid is present in the far IR. Also in contrast to the previous section, here we find ${\displaystyle{ \lim_{r \to \infty}}} \int_{\R^2} \star \left[ Z(\Phi) F \right] \sim r^{-7/3} \to 0$, so that there is no conserved flux emanating from the `horizon' in the far IR. These solutions therefore describe a fully mesonic phase, with all flux sourced by the fermion fluid. In fact, the solution (\ref{eq:f1}) -- (\ref{eq:p1}) can be understood in the following simple way. The leading behavior for the metric and dilaton, i.e. $\{f,g,\Phi\}$, is precisely that of the `domain wall' spacetime with no electric flux or fluid, i.e. $h=\hat \sigma = 0$. We proceed to sprinkle a density of charged fluid onto this domain wall spacetime. These fermions then self-consistently source an electric field according to the Gauss law in the domain wall background, giving at leading order in the IR the expression that we have already quoted
\be
\frac{21}{16} h_0 = \hat \sigma(h_0) \,.
\ee
Because $Z \to 0$ in the far IR, we might expect that the backreaction of the electric flux and fermion fluid on the spacetime and dilaton will be subleading as $r \to \infty$. Indeed, this is what occurs.

As in the previous subsection, the far IR solution is not scale invariant and suffers from divergent curvature invariants leading to a null singularity. In the present case, the `string coupling' is also divergent in the IR, so string loop effects will become important at the lowest energy scales.

Turning on a very small temperature, it is plausible that the specific heat will be that of the finite temperature domain wall solution, with the effects of the electric flux subleading in the IR, as we found for the zero temperature solution. That problem was studied in some generality in, for instance, \cite{Charmousis:2010zz}. For our choice of potential $V$ in (\ref{eq:ZV}), their results show that any nonzero temperature leads to a black hole solution, with specific heat $c_s \sim T^{3/2}$ at small temperatures. The system is therefore again gapless, although there are fewer degrees of freedom at the lowest energy scales than we found in the fractionalized phase.

By a slightly different choice of potential $V$ in (\ref{eq:ZV}) we can easily find a model in which the domain wall spacetime, and hence presumably the mesonic phase we are presently discussing, is gapped. In this case a black hole does not form until the system is heated above a deconfinement temperature $T_C$, see e.g. \cite{Charmousis:2010zz}. While it is natural to associate fractionalized phases with deconfinement, the two are logically distinct. In the model we have chosen to focus on, the `glue' remains deconfined throughout, and yet the system exhibits a fractionalization transition in the charged sector. The situation is similar to transitions in some holographic models of QCD, wherein the quarks remain bound into mesons above the deconfinement temperature, and only `melt' at a higher temperature \cite{Mateos:2006nu}.

\section{The full solutions and phase diagram}

Putting together all the ingredients above we can map out the zero temperature phase diagram of our model.
In the UV theory we have two relevant couplings, the chemical potential $\hat \mu$ and the coupling $\phi_0$ for the operator $\ocal$ dual to the dilaton in the bulk. We can therefore characterize the theory as a function of the dimensionless ratio $\phi_0/\hat \mu$.
In the following we present results for the free energy $\hat \Omega$, charge density $\hat Q$ and the ratio of fractionalized to total charge $\hat Q_\text{frac.}/\hat Q$ as a function of $\phi_0/\hat \mu$.

The equations (\ref{eq:ff}) -- (\ref{eq:flux}) are solved by starting in the far IR with the metrics of sections \ref{sec:a} or \ref{sec:b}. The only free parameter is the strength of the irrelevant perturbation about the IR solutions. For a given value of this parameter we numerically integrate the solutions up to the asymptotic region at small values of $r$. Near the boundary we can read off the ratio $\phi_0/\hat \mu$ as well as thermodynamic quantities, by fitting to the boundary expansions of section \ref{sec:ads}. In performing the integration we must set the fluid variables $\{\hat \sigma, \hat \rho, \hat p \}$ to zero whenever $\hat \mu_{\text{loc.}}(r) < \hat m$ and keep them in the equations otherwise.

\begin{figure}[h]
\begin{center}
\includegraphics[height=130pt]{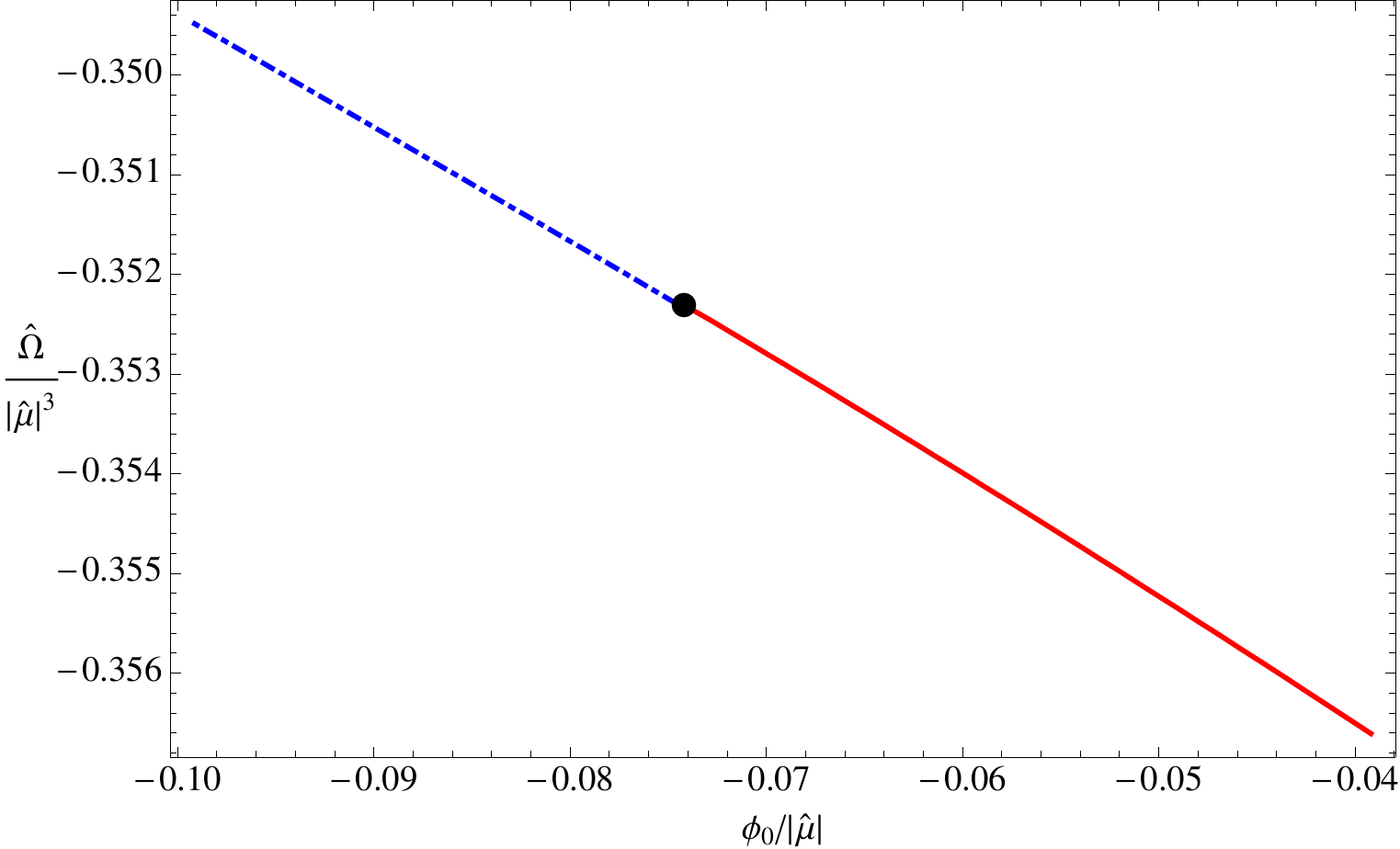}\includegraphics[height=132pt]{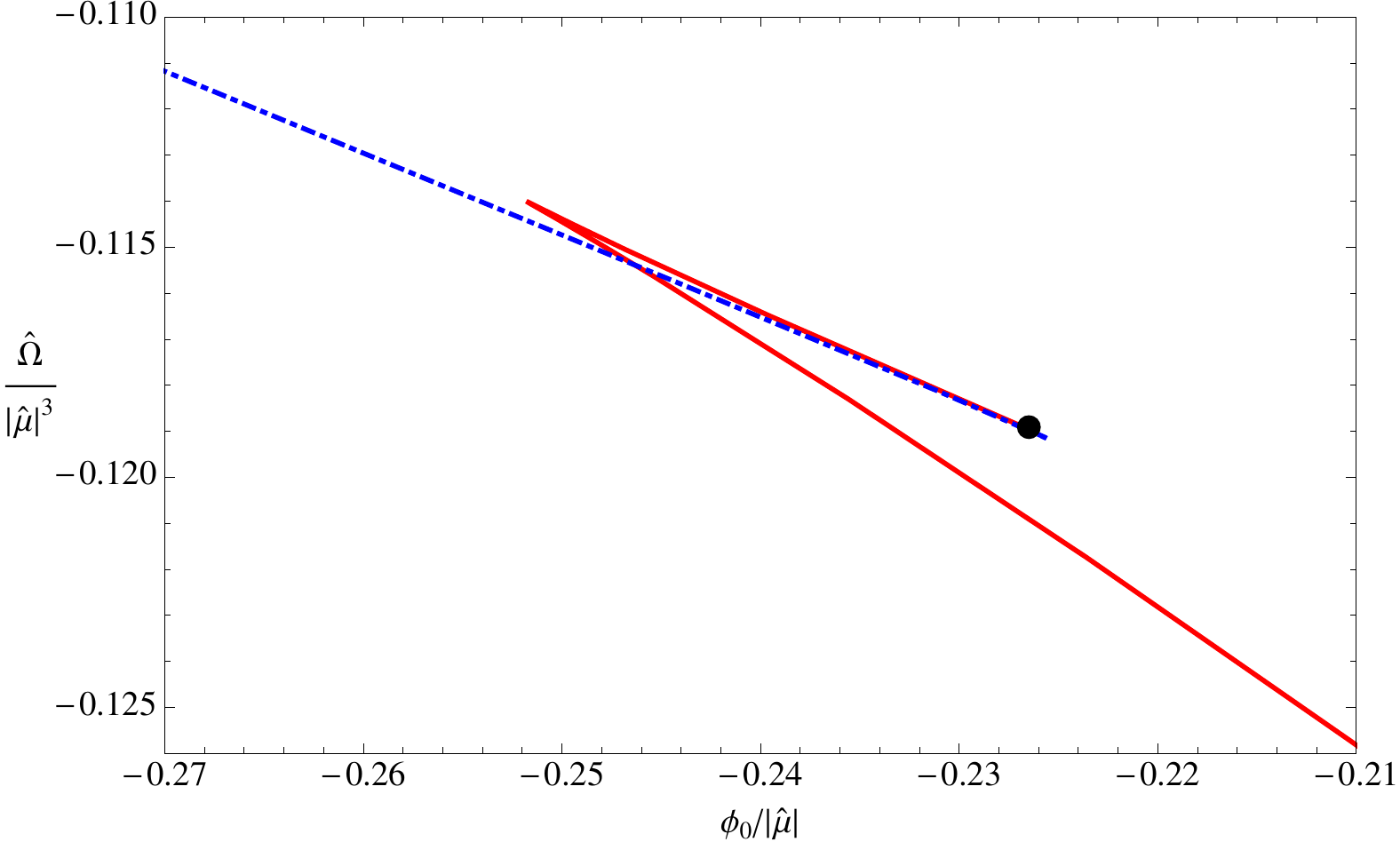}
\caption{Free energy $\hat \Omega$ as a function of the relevant coupling $\phi_0$ for $\{\hat m, \hat \beta\} =\{0.1, 20 \}$ (left) and
$\{\hat m, \hat \beta\} = \{0.5, 10 \}$ (right).
Dashed blue lines indicate a mesonic phase while the solid red lines indicate a partially fractionalized phase. The black dot denotes the location of the Lifshitz fixed point. The left plot exhibits a continuous fractionalization transition, while the right plot exhibits a first order transition.
 \label{fig:omega}}
\end{center}
\end{figure}

In figure \ref{fig:omega} below we plot the free energy as a function of the relevant coupling $\phi_0$ at fixed chemical potential $\hat \mu$. We always plot dimensionless ratios. Two cases are plotted, corresponding to different values of the fermion fluid parameters $\{\hat m, \hat \beta\}$. The
properties of the Lifshitz fixed point for the two values we chose are shown in the following table. These values are
illustrative of continuous and first order fractionalization transitions
\begin{table}[h]
\begin{center}
\begin{tabular}{|c|c|c|c|c|}
\hline
$\{\hat m, \hat \b \}$ & $z$ & $ [g_\text{irrel.}]$ & $ [g_\text{rel.}]$ & stability \\
\hline
$\{0.5, 10 \}$ & $2.62$ & $-2.43$ & $2.31 \pm 1.25 \, i$ & unstable \\
\hline
$\{0.1, 20 \}$ &  $1.42$ & $-1.11$ & $1.52$ & stable \\
\hline
\end{tabular}
\caption{An example of a stable and unstable Lifshitz fixed point.}
\end{center}
\end{table}%
and can also be read off from the plots in section \ref{sec:lif} above. As we see in the free energy plots of figure \ref{fig:omega}, when the relevant operator about the Lifshitz fixed point has a complex scaling dimension, then the Lifshitz point is in fact a local maximum rather than a minimum of the free energy. The continuous fractionalization transition in this case occurs on an unstable branch of solutions, and so the continuous transition is preempted by a first order transition, as the two stable branches (the minima) cross.
In these plots and throughout this section, we use dashed blue lines to denote mesonic phases and solid red lines to denote phases with partial or total fractionalization.
Points on the blue lines are obtained by integrating outwards from the near horizon geometry of section \ref{sec:b} while points on the red lines are obtained by integrating out from the near horizon geometry of section \ref{sec:a}. As an independent check of the numerics, we also integrate out from the Lifshitz solution. The plots in figure \ref{fig:omega} are very zoomed in to the critical point; it is necessary to use high precision numerical integration and to read off the asymptotic quantities with some care.

Combining the information in figures \ref{fig:zcontour} and \ref{fig:reimcontour} we can conclude that continuous fractionalization transitions, with real scaling dimension for the relevant operator at the Lifshitz fixed point, can only occur for either relatively small or relatively large dynamical critical exponents $z$. More precisely, we only find continuous transitions for $z \lesssim 1.46$ or $z \gtrsim 6.98$. The bounds are attained in the case of vanishingly small mass.

\begin{figure}[h]
\begin{center}
\includegraphics[height=130pt]{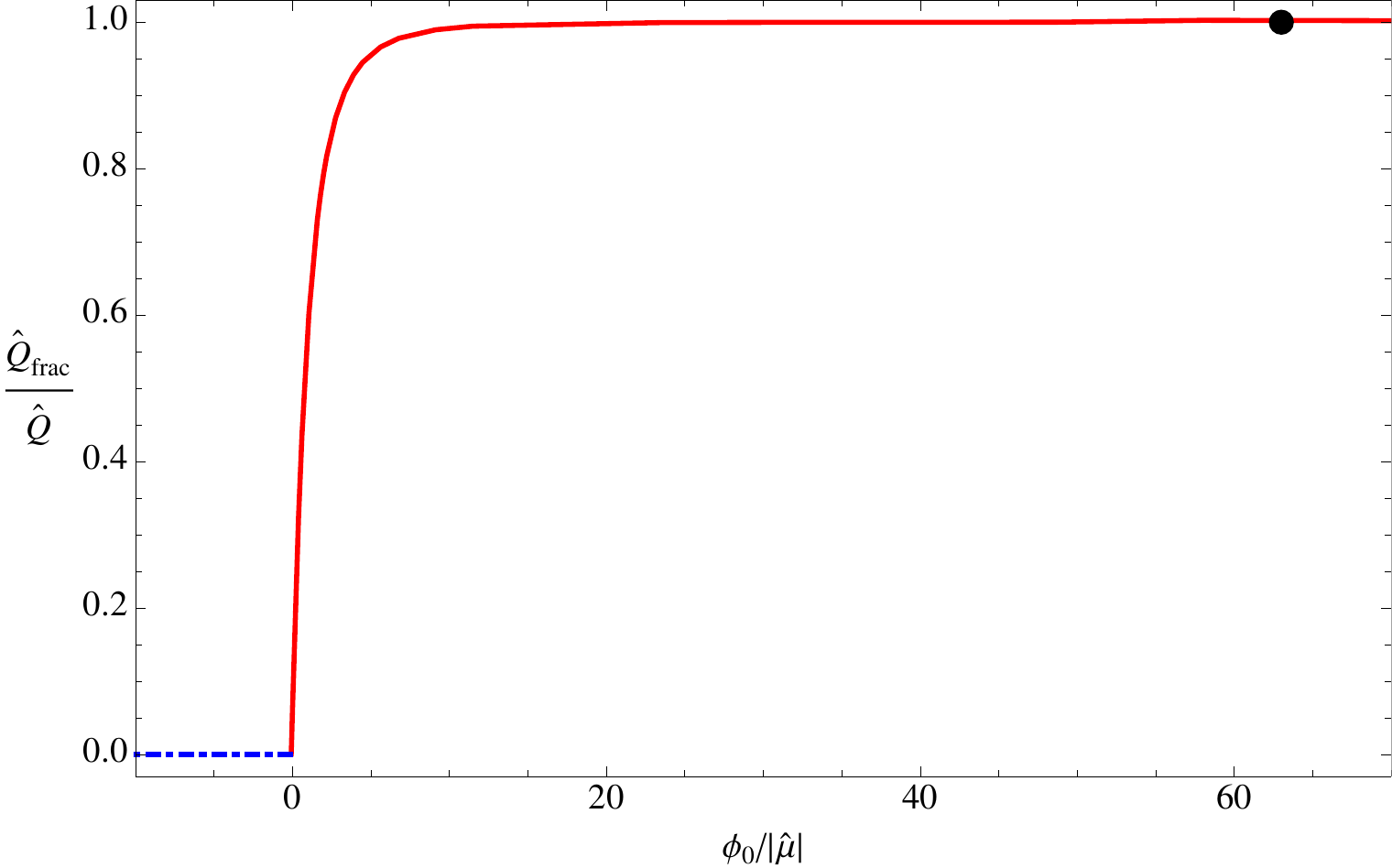}\includegraphics[height=130pt]{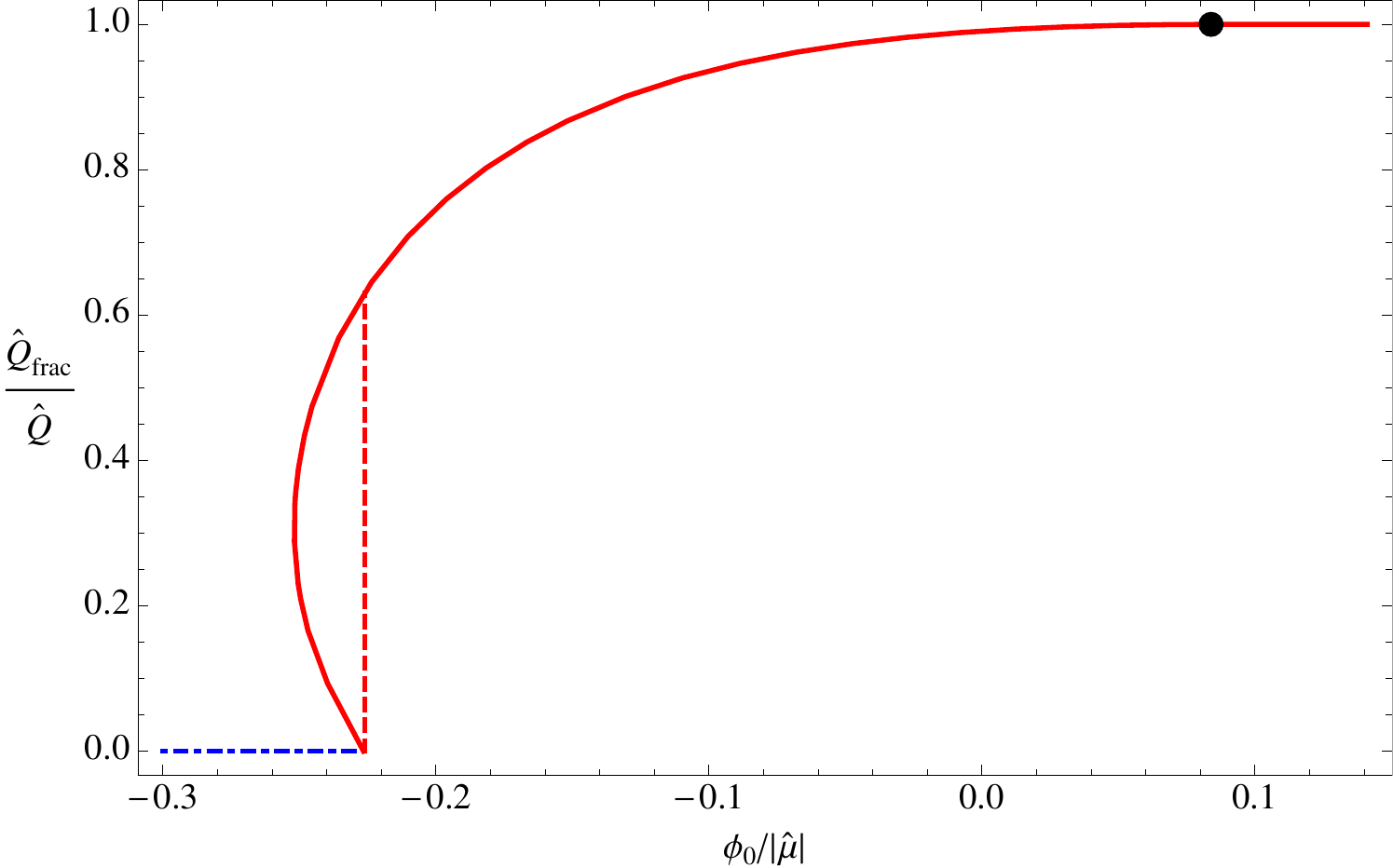}
\caption{Ratio of fractionalized charge to total charge $\hat Q_\text{frac.}/\hat Q$ as a function of the relevant coupling $\phi_0$ for $\{\hat m, \hat \beta\} = \{0.1, 20 \}$ (left) and $\{\hat m, \hat \beta\} = \{0.5, 10 \}$ (right).
The dashed blue lines indicate a mesonic phase while the solid red lines indicate partially and fully fractionalized phases. The black dot denotes the location of the third order transition between partial and full fractionalization. The dashed line in the right hand plot indicates the jump in the ratio at the first order transition. \label{fig:ratio}}
\end{center}
\end{figure}

By integrating the Gauss law (\ref{eq:flux}) we obtain the charge density as
\be
\hat Q = \int_{r_1}^{r_2} \frac{\sqrt{g}}{r^2} \hat \sigma dr + \frac{1}{V} \lim_{r \to 0} \int_{\R^2} \star \left[Z(\Phi) F \right] \equiv \hat Q_\text{mes.} +  \hat Q_\text{frac.} \,.
\ee
Here $r_1$ and $r_2$ are the radii between which a nonzero density of the charged fluid is present. These are the radii for which $\hat m < \hat \mu_\text{loc.}(r)$. For the mesonic phase $r_2 \to \infty$, as the charged fluid extends all the way into the IR. For fully fractionalized phases, in contrast, there are no radii for which the fluid is present. In general, given a numerical solution, we can read off the ratio $\hat Q_\text{frac.}/\hat Q$ of fractionalized to total charge density. This ratio is plotted in figure \ref{fig:ratio} below as a function of the relevant UV coupling $\phi_0$ for the same values of $\{\hat \beta, \hat m\}$ as we used in the previous free energy plots.

The plots exhibit two transitions, between mesonic and fractionalized and between partial and full fractionalization.
As is apparent in the plots, the latter phase transition is softer than the former. The softer transition occurs as the width $\Delta r = r_2 - r_1$ occupied by the fermion fluid vanishes. The transition is very similar to that which occurs upon heating up the electron star without the dilatonic deformation \cite{Hartnoll:2010ik, Puletti:2010de}. That transition was found to be third order. We refer the reader to those papers for more details, as this transition is not our main concern here. The third order nature of the transition can be heuristically understood as follows \cite{Hartnoll:2010ik}. The difference in free energy due to the presence of the fluid is just the pressure of the fluid integrated over the radii at which it is present: $\Delta \hat \Omega \sim \Delta r \cdot \hat p \sim \Delta r \cdot \delta \hat \mu_\text{loc.}^{5/2}$. Here we defined $\delta \hat \mu_\text{loc.} = \hat \mu_\text{loc.} - \hat m$ and used the expression (\ref{eq:eos}) for the pressure of the fermion fluid when the local chemical potential is only slightly bigger than the mass, as pertains close to the critical point where the fluid is about to disappear. Close to the transition one finds that $\Delta r \sim \delta \hat \mu_\text{loc.}^{1/2} \sim |\hat \mu - \hat \mu_C|^{1/2}$. Putting together the above formulae gives the third order transition $\Delta \hat \Omega \sim |\hat \mu - \hat \mu_C|^3$.

By varying $\{\hat \beta, \hat m \}$, it may be possible to achieve a first order transition directly from a mesonic to a fully fractionalized phase. This seems most likely for large masses. For masses bigger than the critical mass of figure \ref{fig:mmax} there is no Lifshitz solution to mediate between the mesonic and fractionalized phases. The transition is then necessarily first order.

While $\hat Q_\text{frac.}/\hat Q < 1$ is a well defined bulk measure of the presence of mesino charge carriers in the bulk, it does not provide a useful field theory characterization of this fact. Two field theory descriptions of the same information can be obtained either by integrating the occupation numbers of the mesino fermion operator bilinear \cite{vCubrovic:2010bf, Hartnoll:2010ik}
\be\label{eq:integral}
n = \int \frac{d\w d^2k}{(2\pi)^3}\langle \Psi^\dagger_{\w,k} \Psi_{\w,k} \rangle \,,
\ee
or by summing up the volume of all the 2+1 dimensional Fermi surfaces in the boundary mesino correlators \cite{Hartnoll:2011dm, Iqbal:2011in, Sachdev:2011ze}
\be\label{eq:sum}
\hat Q - \hat Q_\text{frac.} = \sum_n \frac{2}{(2\pi)^2} V_n \,.
\ee
The bulk WKB limit, that we have taken in order to use a fluid description for the fermions, implies that there are many Fermi surfaces in the sum (\ref{eq:sum}). We can think of these as different excited mesino states. The WKB limit also implies that the tails of the fermion wavefunctions go to zero very quickly outside the boundary of the fluid, and so can be neglected for many purposes. Nonetheless, it is from these tails that we would be able to extract the expectation value in (\ref{eq:integral}).

\begin{figure}[h]
\begin{center}
\includegraphics[height=200pt]{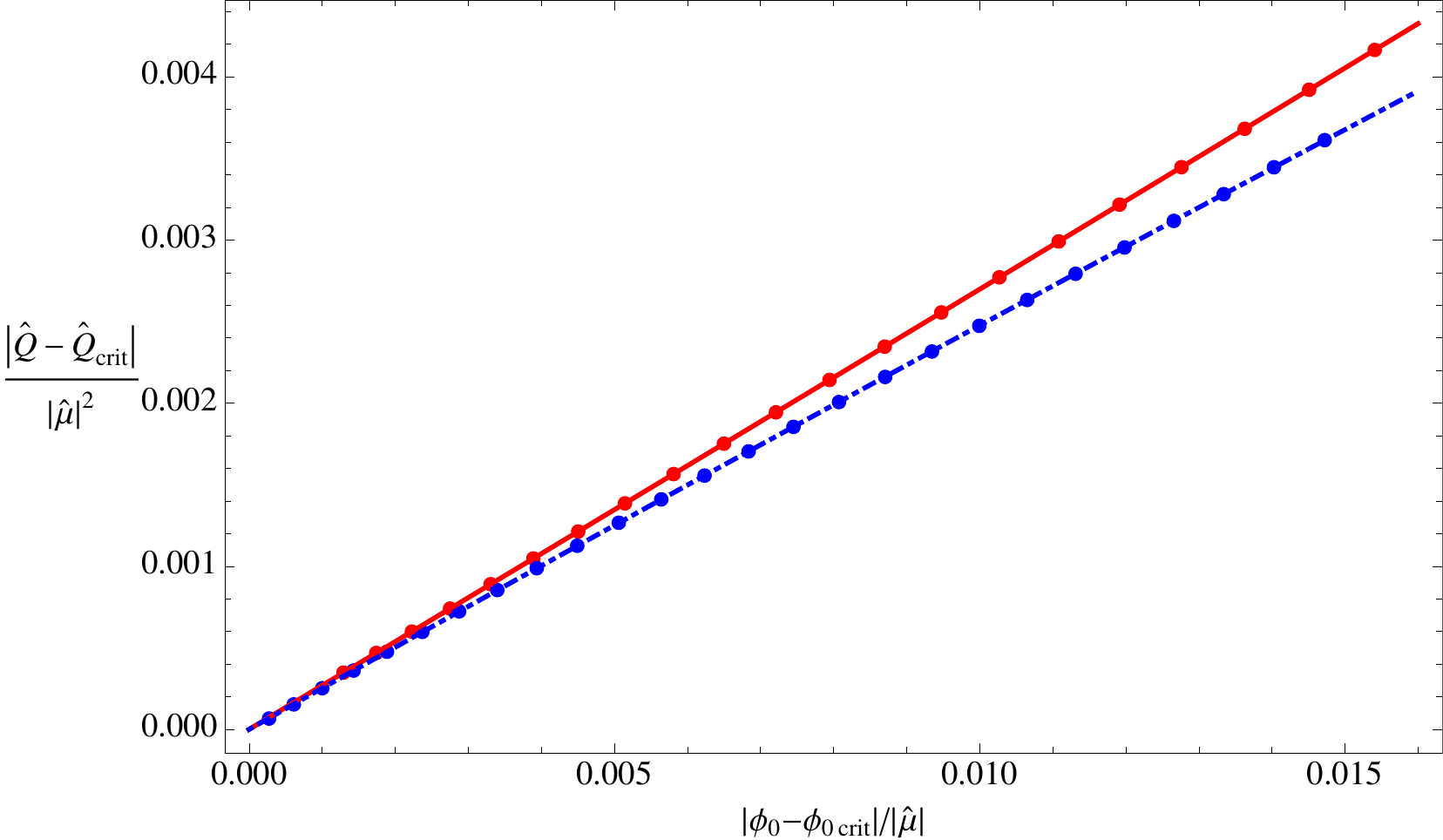}
\caption{Jump in the gradient of the charge density across the continuous fractionalization transition. Data points together with fit.
\label{fig:charge}}
\end{center}
\end{figure}

Finally, let us return to the continuous fractionalization transition. The behavior of the ratio of fractionalized charge in figure \ref{fig:ratio} is suggestive of a second order phase transition, despite not being a symmetry breaking order parameter. To verify that this is the case, we can ask whether the charge susceptibility $\pa^2 \hat \Omega/\pa \hat \mu^2$ is discontinuous across the transition. This is equivalent to the gradient of the charge density $\hat Q \sim \pa \hat \Omega/\pa \hat \mu$ jumping across the transition. The charge density can be extracted quite robustly from our numerics, as can, using the critical Lifshitz solution, the critical values of the ratios $\hat Q/\hat \mu^2$ and $\phi_0/\hat \mu$. Nonetheless, the question turns out to be delicate. In figure \ref{fig:charge} below we plot data points very close to the phase transition, together with fits of the form $y = a x + b x^2$.
The slope $a$ in the fit jumps from $\approx 0.269$ on one side of the transition to $\approx 0.253$ on the other. This is a change of about $6\%$. We have tried to ensure that our numerics are sufficiently accurate to pick out this small jump and hence conclude that the transition is indeed second order. We hope in the future it will be possible to rule out a higher order transition analytically.

\section{Final remarks}

The primary result of this paper is the construction of a simple holographic framework where the fraction of the charge density associated to
gauge invariant Fermi surfaces could be tuned as a function of a relevant coupling in a quantum field theory. We exhibited phase transitions between `mesonic' phases, wherein the Fermi surfaces account for all of the charge, to partially and fully fractionalized phases where they do not. At least for our models, this shows that neither the presence nor absence of gauge invariant Fermi surfaces is generic in holographic theories, it depends on relevant couplings in the theory.

The phases we found broadly correspond with those anticipated in \cite{Huijse:2011hp} as FL, FL$^*$ and NFL phases.
A feature of the particular model that we focused on is that it distinguished deconfinement of the glue sector from fractionalization of the charged sector. Our theories were gapless in all three phases, fractionalized or not. The class of models we have considered also admit gapped phases, and it is certainly of interest to study the interplay of confinement and Higgsing with fractionalization, in order to exhibit phase transitions into truly Fermi liquid phases \cite{Huijse:2011hp, Sachdev:2011ze}.

For convenience we have worked in a coarse-grained fluid limit. This limit has the feature that there are many bulk Fermi surfaces. It has recently been demonstrated that it is technically feasible to go beyond this limit \cite{Sachdev:2011ze}, and it will be important to generalize our discussion in that direction. That said, given that we start with a fractionalized, i.e. gauge-theoretic, description of the theory at large $N$, it is perhaps not surprising that once a single bound state forms, it is easy to form many further bound states. Whatever is the correct theory of quantum gravity for the universe in which we actually live, it must be compatible with the fact that we are surrounded by fluids. The fluid limit is natural in the bulk description and may yet have a deeper role to play in holography. This observation was also one of the motivations behind the earlier paper \cite{Arsiwalla:2010bt}.

There is likely to be a bosonic analogue of the phase transitions we have discussed. In that case the mesonic phase will be superconducting rather than composed of Fermi surfaces \cite{Hartnoll:2008kx}, and the onset of complete fractionalization will be accompanied by the recovery of the global $U(1)$ symmetry. In fact, interesting phase transitions with this flavor have already been obtained in an Einstein-Maxwell-dilaton-charged scalar theory in \cite{Gauntlett:2009bh}.

Throughout our phase diagram, the geometries become singular in the far interior, as we have discussed above. All of these singularities are well behaved in the sense that they can be regulated by heating up the system by a very small amount. Nonetheless, the singularities indicate that the true far IR of the system at zero temperature may revel new physics. In particular, it is possible that in the fractionalized phases e.g. higher derivative corrections will stabilize the dilaton at a highly curved $AdS_2 \times \R^2$ IR fixed point, along the lines of \cite{Sen:2005wa}, or alternatively gap the theory.

The fractionalization transitions that we have found do not realize one of the key features desired in \cite{FFL3}. Namely, the fractionalization proceeds by the Fermi surfaces successively vanishing as their Fermi momenta $k_F \to 0$, as the local bulk chemical potential drops below the fermion mass, in contrast to the residues vanishing, $Z \to 0$, with the Fermi surface size remaining finite. While it is possible to obtain holographic Fermi surfaces with vanishing residue given an $AdS_2 \times \R^2$ near horizon geometry \cite{Faulkner:2009wj}, those Fermi surfaces are not in fact themselves fractionalized (although they do interact with a fractionalized sector described by the extremal horizon), precisely because they appear as singularities in gauge invariant mesino Green's functions. To obtain a holographic fractionalization transition in the spirit of \cite{FFL3} one must take the theory into a Higgsed phase.

Finally, it is of interest to define an order parameter, a generalization perhaps of the Polyakov loop, that is capable of distinguishing fractionalized and mesonic phases. Our phase transitions give a test bed in which to construct and understand such an order parameter holographically. Once defined as a field theoretical quantity, such an order parameter should be useful beyond holography.

\section*{Acknowledgements}

We would especially like to acknowledge many helpful conversations with Diego Hofman, Subir Sachdev and Alireza Tavanfar on topics closely related to this project.
In the final stages of the project we benefited greatly from the stimulating environment of the KITP in Santa Barbara, during the workshop on 
Holographic Duality and Condensed Matter Physics, and in particular from discussions with Matthew Fisher, Chris Herzog, Nabil Iqbal, Matthias Kaminski, John McGreevy, Todadri Senthil and Julian Sonner.
This research was supported in part by the National Science Foundation under
Grant No. NSF PHY05-51164. The work of S.A.H. is partially supported by a Sloan research fellowship. L.H. acknowledges funding from the Netherlands Organisation for Scientific Research (NWO).

\end{document}